\documentstyle[prc,twocolumn,aps,epsfig]{revtex}
\preprint{NT@UW-99-10new version}
\begin{document}
\draft
\title{QCD Rescattering and High Energy Two-Body Photodisintegration of 
the Deuteron}
\author{Leonid L. Frankfurt$^{a}$, Gerald A. Miller$^{b}$,
Misak M. Sargsian$^{b,c*}$ and Mark I. Strikman$^{d}$}
\address{
$^{a}$School of Physics and Astronomy,
Tel Aviv University, Tel Aviv 79978, Israel \\
$^{b}$Department of Physics, University of Washington, Box 351560,
Seattle, WA 98195-1560, USA \\
$^{c}$Yerevan Physics Institute, Yerevan 375036, Armenia\\
$^{d}$Department of Physics, Pennsylvania State University,
  University Park, PA 16802, USA\\
}
\date{\today{}}
\maketitle
\narrowtext
\begin{abstract}
Photon absorption by a quark in one nucleon followed by its  high momentum
transfer interaction with a quark in the other may produce two final-state
nucleons with high relative momentum. We sum the relevant quark rescattering 
diagrams, to show that the scattering amplitude depends on a convolution
between the large angle $pn$ scattering amplitude, the hard photon-quark
interaction vertex and the low-momentum deuteron wave function. The computed
absolute values of the cross section are in reasonable agreement with the data.
\end{abstract}
\pacs{}

\vspace{-0.4cm}
\narrowtext
The recent experiments on high energy two-body photodisintegration of the
deuteron\cite{E89012,NE8,NE17}, set a new stage in high energy
(E$_\gamma \ge$ 1 GeV) nuclear physics. The calculations using the
conventional mesonic picture of nuclear interactions failed to describe
the qualitative features of  these measurements. See Ref.\cite{E89012} for a
comparison between the different approaches. Thus these experiments are
unique in testing the implications of quantum chromodynamics QCD in nuclear
reactions~\cite{BCh,Holt}.

The hypothesis that the Fock state with the smallest number of partonic 
constituents dominates in the two body large angle hard collisions
predicts with considerable success the scaling behavior of the fixed angle
differential cross sections\cite{hex}. In particular, the cross section of
the fixed angle $\gamma d \rightarrow pn$ reaction should scale as
${d\sigma/dt\sim s^{-11}}$. This scaling idea is successful in describing
a number of processes; but it leads to many unanswered questions
\cite{Isgur_Smith,Rady}.
In particular what suppresses the
contributions  of components of the hadronic wave function consisting of
a single fast parton  and wee spectator partons
(Feynman mechanism)\cite{Feynman}? 
In a pQCD analysis of exclusive reactions, one is faced
with the additional problem of  describing the absolute value of the cross
section, (in reactions involving baryons, the calculations
underestimate   measured cross sections by orders of magnitude 
see e.g. \cite{Isgur_Smith}). Examining  the data \cite{E89012} shows
that the situation is rather complicated for $\gamma d \rightarrow pn$
reactions too. Scaling is observed  at  $\theta_{cm}\approx 89^0,69^0$, but
not for smaller angles. No existing model is able
to calculate the absolute value of the cross section\cite{E89012}.

Here we investigate the effects in which the absorption of the photon
by a quark of one  nucleon, followed by  a high-momentum transfer
(hard) rescattering with a quark from the second nucleon, produces the 
final two nucleon state of large relative momenta.
We calculate the contribution which comes from the low relative momentum
($<$ 300 MeV/c) of two nucleons. This contribution is dominant since the
deuteron is a loosely bound system. Another contribution\cite{LN}, occurring 
when the photon breaks up a pre-existing high relative momentum state in the 
deuteron depends on the deuteron wave function  evaluated at relative nucleon 
momenta $\sim$ few~GeV/c.

The validity of our hard-rescattering mechanism requires  
several kinematic conditions. The use of the partonic picture
requires that the masses of the intermediate hadronic state produced by the 
$\gamma N$ interaction  be larger than some minimum mass characterizing the
threshold to reach the continuum. This is known from deep inelastic
scattering\cite{Feynman} to be $W\approx 2.2~GeV$. Here  the mass of
the intermediate (between the photon absorption and quark rescattering)
virtual state is $m_{int}\sim \sqrt{2E_\gamma m_N}$. {}From the condition
$m_{int}\ge W$ one obtains $E_{\gamma}\gtrsim 2.5~GeV$.
Next, the struck quark (Fig.~1) should be energetic enough to reach the
quark of the other nucleon without radiating soft (bremsstrahlung) gluons.
Such radiation is characterized by a regeneration distance 
$l_{r}~\approx ~E_\gamma\cdot R^2$\cite{basicsPQCD}, where
$R\approx 0.3$ fm  characterizes the confinement radius.
For $E_\gamma \ge 2.5$ we obtain $l_{r}  \gtrsim 1.1 $ fm  
is larger than the nucleon radius and comparable with relevant 
internucleon distances in deuteron. 
Finally, to ensure that the quark
rescattering is hard enough, one requires that the transverse momentum of
final nucleons $p_{t}\ge 1$ GeV which requires $-t,-u\gtrsim 2 $ GeV$^2$.

Our derivation proceeds by evaluating Feynman diagrams such as Fig.~1.
The quark interchange mechanism (see e.g.\cite{GBB}), in which quarks are 
exchanged between nucleons via the exchange of a gluon 
and the recoil quark-gluon system is on mass shell, is used. 
All other quark-interactions are included in the  non-perturbative partonic 
wave function of the nucleon, $\psi_N$.

One proceeds (see e.g. \cite{BL,FS}) by integrating over the
minus-momenta\cite{lc} to obtain an expression involving only
three-dimensional integrals. The result is that
the deuteron-NN vertices are replaced by
the light-cone deuteron wave  function, $\Psi_d$ and nucleon-(quark,gluon)
vertices are replaced by   $\psi_N$.  
We use  a simplified notation in which only the momenta of the
interacting quarks are labelled. 
Thus the scattering amplitude $T$ for photo-disintegration of a deuteron
($p_d$, mass $M_d$) into two nucleons ($p_A$ and $p_B$) is:   
\vspace{-0.23cm}
\begin{eqnarray}
&& T = -\sum\limits_{e_q} \int  \left( 
{\psi_N^\dag(x'_2,p_{B\perp},k_{2\perp})\over x'_2}\bar u(p_B-p_2+k_2)\right. 
\nonumber \\
&&\left[-igT_c^{F}\gamma^{\nu}\right]
{u(k_1+q)\bar u(k_1+q)\over (k_1+q)^2 - m_q^2 + i\epsilon}
\left[-ie_q\epsilon^{\perp}\cdot\gamma^{\perp}\right]\nonumber \\
&&\left. u(k_1)
{\psi_N(x_1,p_{1\perp},k_{1\perp})\over x_1}\right)\left\{
{\psi_N^\dag(x'_1,p_{A\perp},k_{1\perp})\over x'_1}\right.\nonumber \\
&&\bar u(p_A-p_1+k_1)\left[-igT_c^{F}\gamma_{\mu}\right]u(k_2) 
\left. {\psi_N(x_2,p_{2\perp},k_{2\perp})\over x_2}\right\}
\nonumber \\ 
&& G^{\mu\nu} {\Psi_{d}(\alpha,p_{\perp})\over 1-\alpha}
{dx_1\over 1-x_1} {d^2k_{1\perp}\over 2(2\pi)^3}
{dx_2\over 1-x_2} {d^2k_{2\perp}\over 2(2\pi)^3}
{d\alpha\over \alpha} {d^2p_\perp\over 2(2\pi)^3}. \! \! \!
\label{Ta}
\end{eqnarray}
The deuteron is composed of nucleons of momenta $p_1$ and  $p_2$  with
$\alpha \equiv {p_{1+}\over p_{d+}}$, $p_2=p_d-p_1$ and $p_{1\perp}=
-p_{2\perp}\equiv p_\perp$\cite{lc}.
Each of these consists of one active  quark of momenta $k_1$ and $k_2$ 
and a residual quark-gluon spectator system of momenta 
$p_1-k_1$ and $p_2-k_2$. It is useful to define the momentum fractions:
$x_1\equiv {k_{1+}\over p_{1+}} = {k_{1+}\over \alpha p_{d+}}$, 
$x_2\equiv {k_{2+}\over p_{2+}} = {k_{2+}\over (1-\alpha) p_{d+}}$,
$1-x'_1 = {p_{1+}-k_{1+}\over p_{F+}}$  and
$1-x'_2\equiv {p_{2+}-k_{2+}\over p_{B+}}$.
The amplitude in Eq.~(\ref{Ta}) is a convolution of several  blocks.
a) $\Psi_{D}(\alpha, p_{2\perp})$ describes the transition of
the deuteron into a  two-nucleon system.
b) The term in the $\left(...\right)$ describes the ``knocking out'' of the
quark of one nucleon by the incoming photon, 
with intact quark-gluon recoil  of interacting nucleon
and subsequent gluon exchange
of that quark with the quark of the second nucleon. It consists of $\psi_N$;
the $\gamma$-quark vertex $-ie_q\epsilon^\perp\cdot\gamma^\perp$, where 
$\epsilon^\perp$ is polarization vector of incoming photon,  the
intermediate-state propagator of the knocked-out quark 
${u\bar u\over (k_1+q)^2-m_q^2+i\epsilon}$\cite{contact}, with
current quark mass $m_q$; the quark-gluon vertex $=igT_c^F\gamma^\mu$;
and the wave function of the final nucleon $ \psi_N^\dag$. 
c) The expression in $\left\{...\right\}$ describes the interaction of the
quark from second nucleon with the knocked-out quark;
d) The propagator of the exchanged gluon is
$G^{\mu\nu}= {id_{\mu\nu}\over
\left[l - q + (p_1- k_1) - (p_2-k_2) \right]^2 + i\epsilon}$
with polarization matrix $d_{\mu\nu}$ (fixed using the light-cone gauge)
and $l\equiv (p_B-p_1)$.
We use the reference frame where $p_d = (p_{d0},p_{dz},p_{\perp})\equiv
({\sqrt{s'}\over 2}+{M_d^2\over 2\sqrt{s'}},
{\sqrt{s'}\over 2}-{M_d^2\over 2\sqrt{s'}} ,0)$,
with  $s = (q+p_d)^2$, $s'\equiv s-M_D^2,$ and 
$q^\mu = ({\sqrt{s'}\over 2}, -{\sqrt{s'}\over 2},0_\perp)$.
\vspace{-0.4cm}
\begin{figure}[h]
\begin{center}
\epsfig{angle=0,width=2.8in,file=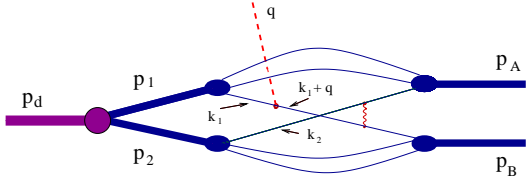}  
\caption{Quark Rescattering diagram.}
\end{center}
\label{Fig.1}
\end{figure}
\vspace{-0.42cm}
To proceed  we analyze  the denominator of the knocked - out quark propagator
when recoil quark-gluon system is on mass shell.
We are concerned with  momenta such that $p^2_\perp\ll m^2_N\ll s'$ 
and $\alpha \sim {1\over 2}$ so we neglect terms of order
$p_\perp^2,m^2_N/s'\ll 1$ to obtain:
\vspace{-0.12cm}
\begin{equation} 
(k_1+q)^2-m_q^2 + i\epsilon \approx x_1s^\prime(\alpha-\alpha_c + i\epsilon),
\label{alphac}
\end{equation}
\vspace{-0.28cm}\\
where $\alpha_c \equiv {x_1 m_R^2+k_{1\perp}^2\over 
(1-x_1)x_1\tilde s}$,
$\tilde s \equiv s'(1+{M_d^2\over s'})$ and
$m_R$ is the recoil mass of the spectator quark-gluon system of the
first nucleon.  
The deuteron wave function  is very strongly
peaked at $\alpha={1\over 2}$ and $p_{\perp}=0$, so the
dominant contribution to $T$ corresponds to $\alpha_c\approx 1/2$. 
According to  Eq.(\ref{alphac})  the integration in Eq.(\ref{Ta}) over
$k_{1\perp}$ in the region
$k_{1\perp}^2\sim {(1-x_1)x_1\tilde s\over 2}\gg x_1m_R^2$ does  provide
$\alpha_c= {1\over 2}$. 
Keeping only the imaginary part  of the quark
propagator (the knocked-out quark is on-mass shell in the intermediate 
state) leads to $\alpha=\alpha_c$ and corresponds to keeping the contribution 
from the soft component of the deuteron wave function. 

Note that produced in the intermediate state quark-
gluon system is typical for the hard processes in DIS
but not for nucleon and even for baryon resonances. It has mass$^2$ 
$\propto s$.
(Off shell quark can not propagate distances as large as 1~fm 
necessary to hit quark of other nucleon and to produce considered process.)
Another consequence of on-mass shellness of  knock-out quark (and recoil 
quark-gluon system)  is that gauge invariance and conservation of e.m. current 
are easily fulfilled.

Next we calculate the photon-quark hard scattering vertex--
$\bar u(k_1+q)[\gamma_\perp]u(k_1)$ and use Eq.~(\ref{alphac}) to integrate
over $\alpha$, by taking into account only second term in the decomposition
of struck quark propagator: $(\alpha-\alpha_c + \epsilon)^{-1} \equiv
{\cal P}(\alpha-\alpha_c)^{-1} - i\pi\delta(\alpha-\alpha_c)$: 
\begin{eqnarray}
&& T =  {i\over 2} \sum\limits_q  
{e_q(\epsilon^++\epsilon^-) 
\over 2 \sqrt{s'}}
\int\left[
\left( {(1-\alpha_c x_1)^2 + {4k^2_{1\perp}\over s'} \over
\alpha_c^2x_1^2+{4k^2_{1\perp}\over s'}}\right)^{1\over 4}
\hspace{-0.2cm} \alpha_c x_1
\right. \nonumber \\
& & \left . \ \ 
+ \left( {\alpha_c^2x_1^2+{4k^2_{1\perp}\over s'}\over
(1-\alpha_c x_1)^2 + {4k^2_{1\perp}\over s'}}\right)^{1\over 4}
(1-\alpha_c x_1) \right]{1\over x_1}    
\nonumber \\
&&{\psi_N^\dag(x'_2,p_{B\perp},k_{2\perp})\over x'_2}
\bar u(p_B-p_2+k_2)\left[-igT_c^{F}\gamma^{\nu}\right]\nonumber \\
&&u(k_1+q){\psi_N(x_1,p_{1\perp},k_{1\perp})\over x_1}
{\psi^\dag_N(x'_1,p_{F\perp},k_{1\perp})\over x'_1}
\nonumber \\
&&\bar u(p_A-p_1+k_1) \left[-igT_c^{F}\gamma_{\mu}\right]u(k_2)
{\psi_N(x_2,p_{2\perp},k_{2\perp})\over x_2} \cdot G^{\mu\nu}
\nonumber \\
&&\times {dx_1\over 1-x_1} {d^2k_{1\perp}\over 2(2\pi)^3}
{dx_2\over 1-x_2} {d^2k_{2\perp}\over 2(2\pi)^3}
{\Psi_{d}(\alpha_c,p_{\perp})\over \alpha_c(1-\alpha_c)}
{d^2p_\perp\over 2(2\pi)^2},
\label{Tb}
\end{eqnarray}
where 
$\epsilon^\pm\equiv{1\over 2}(\epsilon_x\pm i \epsilon_y)$.
The dominant  contribution arising from  the soft
component of the deuteron occurs when $\alpha_c\sim{1\over 2}$.
Thus we  may substitute $\alpha_c={1\over 2}$.
After factoring  out electromagnetic term, one can identify the remaining
integral in Eq.(\ref{Tb}) (up to a scaling factor $f(l^2/s)$) 
as  a particular contribution to the quark interchange mechanism for
the  wide angle nucleon-nucleon scattering amplitude
-$A_{pn}^{QIM}(s,l^2)$. Then summing over the struck quark contributions from
photon scattering off neutron and proton 
\cite{SU(6)} one obtains:
\begin{eqnarray}
 T & \approx &    
{i(e_u+e_d)(\epsilon^++\epsilon^{-}) \over  2 \sqrt{s'}} \nonumber \\
& & \times \int f({l^2\over s})A_{pn}^{QIM}(s,l^2) 
\Psi_{d} ({1\over 2},p_{\perp}){d^2p_\perp\over (2\pi)^2},
\label{Tc}
\end{eqnarray}
where $e_u$ and $e_d$ are the electric charges of $u$ and $d$ quarks.
The factor $f(l^2/s)$ accounts for the difference between  the hard
propagators  in our process and those occurring in wide angle $pn$
scattering. Within the  Feynman mechanism,  the interacting quark carries
the whole momentum of the nucleon  ($x_{1}\rightarrow 1$), thus $f(l^2/s)=1$.
Within the minimal Fock state approximation, the evaluation of the exact
form of $f(l^2/s)$ requires calculating  the sum of $\sim 10^6$
Feynman diagrams in which  five hard-gluon exchanges are distributed between
six quark lines.  Here we use a   qualitative evaluation of
$f(l^2/s)$. The conditions $\alpha_c\sim {1\over 2}$ , $0<x_1<1$
according to  Eq.(\ref{alphac}) require
$k^2_{1\perp}\sim {x_1(1-x_1)s'\over 2}\sim s'$. Thus the dominant
contribution in Eq.({\ref{Tc}) is given by diagrams in which 
the  struck quark exchanges a hard-gluon with  other quarks prior to  the
interaction with the photon (initial state short range $qq$-correlations).
For these diagrams, the interchange of quarks between nucleons in both 
$\gamma d pn$  and $pn$  amplitudes  is  characterized by  the  same 
virtuality in the propagator of gluon exchange between nucleons, estimated
at $\theta_{cm}= 90^0$ to be  $\sim (p_t/3)^2$.
All other propagators that define the short range part of the nucleon wave 
functions are the same.  The other QIM diagrams with final state $qq$
correlations only (which are suppressed in $\gamma dpn$  amplitude) enter
in the $pn$ amplitude with larger virtuality for exchanged gluon between
nucleons $\sim p_t^2$  and will be numerically small compared to the
diagrams with initial state correlations. Thus  no additional combinatorial
factors  enter in the hard  scattering mechanism.
Since the additional factor entering from the electromagnetic vertices
is $\sim 1$,  $f(l^2/s)$ is a function of the order of unity at $90^0$.
In the further analysis presented here we will
use the approximation: $f(l^2/s)=1$.

The amplitude (\ref{Tc}) depends on only small relative momenta of the target
nucleons, therefore we should use a standard non-relativistic (NR) wave
function; according to \cite{FS}: $\Psi_{d}(\alpha,p_{\perp})
= (2\pi)^{3\over 2}\Psi_d^{NR}(p_z,p_\perp)\sqrt{m_N}$.
We compute the differential cross section averaging $|T|^2$ over the spins
of initial photon and deuteron and summing over the spins of the  final
nucleons. One obtains
\begin{eqnarray}
&&{d\sigma^{\gamma d\rightarrow pn}\over dt} = {1\over 16\pi}
{1\over (s-M_d^2)^2}\bar {|T|}^2 
= {\alpha\over 4}\cdot {\pi^3 \over 9  s'^3} \times \nonumber \\
&& \left| \int\sqrt{m_N}
f({l^2\over s})
A_{pn}^{QIM}(s,l^2) \Psi_d^{NR}(p_z=0,p_\perp)
{d^2p_\perp\over (2\pi)^2}\right|^2.
\label{difcrsa}
\end{eqnarray}
$A_{pn}^{QIM}$ represents a sum of all leading-quark interchange diagrams
with rather general structure of spectator quark-gluons system.
Thus $A_{pn}^{QIM}$ absorbs nonperturbative (noncalculable) part of our
calculation. To proceed we observe that the quark interchange topologies
shown to be the dominant contribution for fixed $\theta_{cm}=90^0$ high
momentum transfer (non strange) baryon-baryon and meson-baryon
scattering \cite{h20}. Thus in the region of $\theta_{cm}\approx 90^0$ we
replace $A_{pn}^{QIN}$ by the experimental data - $A_{pn}^{Exp}$.
In this replacement we neglect the  contribution of $u$ channel 
(backward)  rescattering, since it is a numerically small contribution 
to the $pn$ amplitude, and is additionally suppressed due to the presence 
of the electromagnetic vertex of $\gamma-neutron$ scattering.
We also observe  that the integrand in Eq.~(\ref{difcrsa}) is
dominated by small values of $p_\perp \ll p_{A\perp},p_{B\perp}$.
Thus we  evaluate $A_{pn}^{QIM}$ at  $t_N = (p_B-p_d/2)^2$ 
by  pulling this term out of the integral and expressing it through the
differential cross section of $pn\rightarrow pn$ scattering-
${d\sigma^{pn\rightarrow pn}\over dt}$:
\begin{eqnarray}
{d\sigma^{\gamma d\rightarrow pn }\over dt} & = & {4\alpha\over 9}\pi^4\cdot
{1\over s'} C({t_N\over s}){d\sigma^{pn\rightarrow pn}(s,t_N)\over dt}
\nonumber \\ 
& & \times \left| \int\Psi_d^{NR}(p_z=0,p_\perp)\sqrt{m_N}
{d^2p_\perp\over (2\pi)^2}\right|^2.
\label{difcrsb}
\end{eqnarray}
Eq.~(\ref{difcrsb}) shows that the ${d\sigma^{\gamma d\rightarrow pn}\over dt}$
depends on the soft component of the deuteron wave function, the  measured
high momentum transfer $pn\rightarrow pn$ cross section, the scaling factor
$C({t_N\over s})\approx f^2(t_N/s)\approx 1$ at $\theta_{cm}\sim 90^0$
(and  slowly varying as a function of $\theta_{cm}$) 
 and the additional factor coming from the $\gamma-q$ interaction. 
Note that, Eq.(\ref{difcrsb}) is qualitatively different from the Glauber 
approximation. The latter is applicable only when intermediate nucleons 
are near on-mass shell. On the contrary Eq.(\ref{difcrsb}) is derived  
when the mass$^2$ of the intermediate state is ($\sim s\gg m_{N}^2$).
Our  approach is close to that of Ref.~\cite{BH} in which a  nuclear 
amplitude is expressed as a product of a reduced nuclear amplitude 
and nucleon form-factors. Here   the nuclear amplitude is expressed 
in terms of  the $pn$ hard scattering amplitude, and this
allows us to calculate the absolute value of the cross section. 
Note that  the $pn$ cross section scales as $s^{-10}$. This causes  
Eq.(\ref{difcrsb}) to yield the same asymptotic $s^{-11}$ energy 
dependence (at fixed $t/s$)  as  provided by the quark counting rules.

In the numerical calculations we take $C({t_N\over s})=1$.
Our calculations are implemented using the Paris potential model of
$\Psi_d^{NR}$ \cite{Paris}  (but any realistic wave function would
give the same result)  and  the experimental data from 
\cite{npdata1,npdata2} for ${d\sigma^{pn\rightarrow pn}\over dt}$. 
The $pn\rightarrow pn$ data  are not measured at the $t_N$ needed
to evaluate Eq.~(\ref{difcrsb}), so an extrapolation is necessary.
We determine an upper and lower limit for
${d\sigma^{pn\rightarrow pn}\over dt}$ at $t_N$ using the existing
$pn$ data at $t_N^{min,max}$ such that  $-t_N^{min} < -t_N < -t_N^{max}$. 
Then Eq.(\ref{difcrsb}) is computed using both the data at 
$t_N^{min}$ and at $t_N^{max}$, so that the calculation will produce a band. 
Figure 2 shows that calculations are in agreement with the measured
differential cross sections. Moreover the agreement improves for larger
$\theta_{cm}$ which confirms our expectation that $C(t_N/s)\approx 1$
at $\theta_{cm}=90^0$.  The agreement with the data verifies our underlying
hypothesis  that the size  of the photoproduction reaction is determined by
the physics of high-momentum transfer contained in the hard scattering NN
amplitude. The short-distance aspects of the deuteron wave function are
not important. This  hypothesis, if confirmed by additional studies, 
may suggest the existence of new type of quark-hadron ``duality'', where the
sum of the ``infinite'' number of quark interactions could be replaced by the
hard amplitude of $NN$ interaction.

It is worth noting that  the deviations from the calculation based on
the approximation $C(\theta_{cm})=1$ seem to be consistent with $C$
being a function of $\theta_{cm}$ only. Such a dependence may easily
occur in the quark exchange mechanism (see discussion above). In
particular the whole set of available data can  be described by  taking
$C(t_N/s)={-2t_N/s'\over 1+2t_N/s'}\approx {-t/s'\over 1+t/s'}$
including even the data at $\theta_{cm}=36^0$,
where $-t\le 2~GeV^2$ and the hard interaction mechanism can not be applied.
This may indicate that connection between $NN$ and $\gamma d\to pn$ dynamics
extends to a transitional region of $t$. 
More detailed measurements of angular dependence with an extended range
of energies will allow a verification  of such a scaling pattern.
\vspace{-0.4cm}
\begin{figure}[h]
\begin{center}
\epsfig{angle=0,width=3.4in,height=2.4in,file=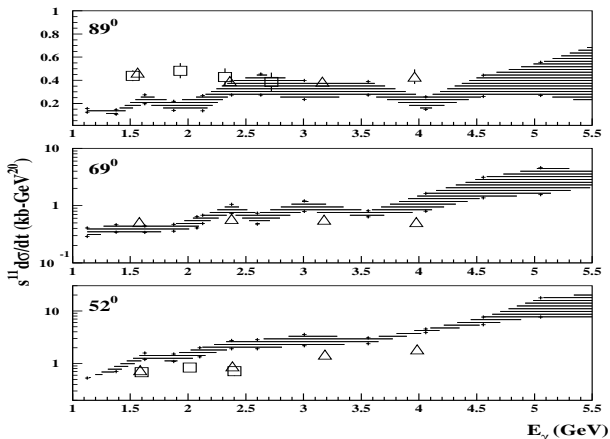}  
\caption{The scaled differential cross sections as a function of the photon
energy, for different values of cm angle. Data are from
[1] (triangles) and [3] (squares).}
\end{center}
\label{Fig.2}
\end{figure}
\vspace{-0.32cm}
One expects also that the photodisintegration reactions would
ultimately be able to address many problems of QCD known from
the hard exclusive two-body reactions. 
More data,  especially with a two proton final state
(i.e. $\gamma+^3He\rightarrow pp$ (high $\ p_t$) + n $(p_t\approx 0$)), and a
more detailed angular distribution would definitely allow such investigations.
The polarization measurement also will be crucial especially at the same $s$
where anomalies observed in hard $pp$ scattering.
Another important extension would be similar experiments using virtual photons.

Although Eq.~(\ref{difcrsb}) is obtained utilizing only the quark interchange
mechanism, the agreement of Fig.2 may suggest that the result is more general
- any hard interaction mechanism for $\gamma d\rightarrow hadrons$ reaction
could be expressed through the contribution to hadron-hadron hard scattering
amplitude. Thus,  the present calculations could be extended to
 reactions with other composition of final high $p_t$ hadrons such as:
($\gamma d \rightarrow \Lambda K N$ or $\Lambda_c D N$). 

The agreement with the recently measured  
cross sections of $\gamma d\rightarrow pn$ reaction 
suggests that such a rescattering mechanism
could be important for any high momentum transfer nuclear reaction. More
detailed data for angular distributions and final states involving different
polarizations and compositions of final hadrons could provide an entirely new
tool for the investigation of QCD dynamics of high energy nuclear reactions.

We thank S.~Brodsky, A.~Dieperink, S.~Nagorny for useful discussions
and R.~Holt and H.~Gao for providing and explaining the experimental data.
This work was supported in part by the U.S. Department of Energy 
and by the Academy of Science of Israel.

$^*$ Present Address: 
Department of Physics, Florida International University, Miami, FL 33199, USA.

\end{document}